\documentclass[aps,prl,showpacs,twocolumn,superscriptaddress]{revtex4-1}
\usepackage[utf8]{inputenc}
\usepackage[american,british]{babel}
\usepackage[T1]{fontenc}
\usepackage[pdftex]{graphicx}  
\usepackage{graphicx, xcolor}
\usepackage{dcolumn}
\usepackage{bm}
\usepackage{amsmath,amsthm,amssymb}
\usepackage{color}
\usepackage{verbatim}
\usepackage{algorithm}
\usepackage[noend]{algpseudocode}

\definecolor{darkGreen}{RGB}{0,110,0}
\definecolor{darkBlue}{RGB}{0,0,130}
\usepackage[colorlinks,citecolor=blue,linkcolor=darkBlue,urlcolor=blue,hyperindex]{hyperref}

\emergencystretch=\maxdimen
\newcommand{\bra}[1]{\langle #1|}
\newcommand{\ket}[1]{|#1\rangle}

\begin{document}

\title{A kaleidoscope of phases in the dipolar Hubbard model}
\author{Tiago Mendes-Santos} 
\email{tmendes@ictp.it}
\affiliation{Instituto de F\'\i sica, Universidade Federal do Rio de Janeiro
Cx.P. 68.528, 21941-972 Rio de Janeiro RJ, Brazil}
\affiliation{The Abdus Salam International Centre for Theoretical Physics, Strada Costiera 11, 34151 Trieste, Italy}
\author{Rubem Mondaini} 
\email{rmondaini@csrc.ac.cn}
\affiliation{Beijing Computational Science Research Center, Beijing, 100193, China}
\author{Thereza Paiva} 
\affiliation{Instituto de F\'\i sica, Universidade Federal do Rio de Janeiro
Cx.P. 68.528, 21941-972 Rio de Janeiro RJ, Brazil}
\author{Raimundo R. \surname{dos Santos}} 
\affiliation{Instituto de F\'\i sica, Universidade Federal do Rio de Janeiro
Cx.P. 68.528, 21941-972 Rio de Janeiro RJ, Brazil}

\begin{abstract}
We investigate the emergence of a myriad of phases in the strong coupling regime of the dipolar   Hubbard model in two dimensions. 
By using a combination of numerically unbiased methods in finite systems with analytical perturbative arguments, we show the versatility that trapped dipolar atoms possess in displaying a wide variety of many-body phases, which can be tuned simply by changing the collective orientation of the atomic dipoles. 
We further investigate the stability of these phases to thermal fluctuations in the strong coupling regime, highlighting that they can be accessed with current techniques employed in cold atoms experiments on optical lattices. 
Interestingly, both quantum and thermal phase transitions are signalled by peaks or discontinuities in local moment-local moment correlations, which have been recently measured in some of these experiments, so that they can be used as probes for the onset of different phases.

\end{abstract}

\date{Version 3.6 -- \today}

\maketitle


\paragraph{Introduction.---}Experiments with ultracold atoms on optical lattices~\cite{Jaksch05,Bloch08,Esslinger10,McKay11} have stimulated the search for new paradigms in many-body physics, especially due to the possibility of controlling and engineering quantum macroscopic states~\cite{Carr09}.
A recent experimental advance is the manipulation of atoms or molecules with (electric or magnetic) dipoles \cite{Lahaye09,Trefzger11,Gadway16}. 
For example, $^{52}$Cr atoms with a large magnetic moment ($6\mu_\mathrm{B}$, with $\mu_\mathrm{B}$ being the Bohr magneton) form Bose-Einstein condensates (BEC's) below $T_c\simeq 700$ nK \cite{Griesmaier05}; 
larger magnetic moments, $\sim\!\! 12\mu_\mathrm{B}$, were later obtained with Er$_2$ molecules~\cite{Frisch15}.
The first quantum degenerate dipolar Fermi gas was realized \cite{Lu12} with $^{161}$Dy atoms cooled down to 20\% of the Fermi temperature, $T_F \approx 300$nK; also, Fermi surface deformation was observed in Er atoms \cite{Aikawa14}.
An ultracold dense gas of fermionic potassium-rubidium ($^{40}$K$-^{87}$Rb) polar molecules was also generated \cite{Ni08}, which paved the way to trap them into 2D and 3D optical lattices \cite{Chotia12}; more recently, a two component  Er dipolar fermionic gas with tunable interactions  was prepared with collisional stability in the strongly interacting regime \cite{Baier18}.

The interest in dipolar atoms stems from the fact that their interactions are long ranged and anisotropic, such that they can be directionally repulsive or attractive. 
This adds extra richness to the diversity of collective states of atoms in an optical lattice \cite{Baranov08,Baranov12}.   
For instance, quantum magnetism of high-spin systems has been experimentally studied with bosonic atoms in optical lattices \cite{dePaz13,Yan13}, and the ability to design quantum spin Hamiltonians with cold atoms may lead to the development of error-resilient qubit encoding and to topologically protected quantum memories \cite{Micheli06}. 
In addition, since one of the motivations to study cold atoms in optical lattices is the possibility of emulating condensed matter models \cite{Jaksch05,Bloch08,Esslinger10,McKay11}, a detailed investigation of effects due to dipolar interactions is clearly of interest.
However, since experimental studies of dipolar fermionic atoms in optical lattices are still in their infancy \cite{Ni08,Chotia12}, theory must take the lead in highlighting interesting effects which would make the experimental effort worthwhile.   
Indeed, several studies suggest that new phenomena may emerge, such as $p$-wave pairing \cite{Bruun08,Cooper09} and different density-wave patterns \cite{Quintanilla09,Lin10,Mikelsons11,He11,Gadsbolle12a,Gadsbolle12b,Bhongale12,Bhongale13,vanLoon15,vanLoon16}, some of which are analyzed through analogies with liquid-crystals.
However, none of these studies predicted the formation of a Mott state, which has recently been achieved in two-component dipolar fermionic systems \cite{Baier18}; thus, a systematic study of the interplay between Mott and competing density-wave patterns is in order.

\begin{figure*}[t]
\includegraphics[scale=0.45]{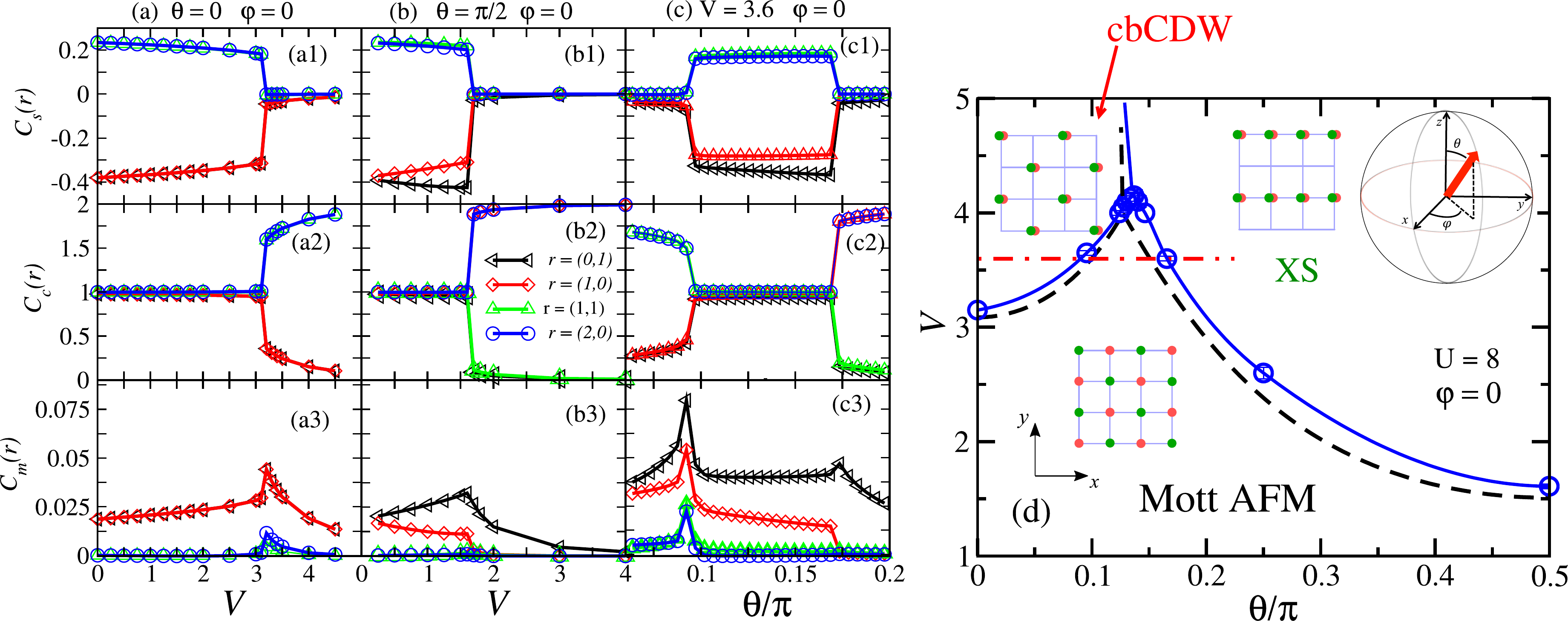} 
\caption{(Color online) 
Panels (a) and (b) show the dependence of the different correlation functions with the dipolar interaction strength, $V$, for dipoles perpendicular (a) and parallel (b) to the lattice; panels (c) show the dependence with the polar angle $\theta$, at fixed $V$.  
Each curve is for a fixed distance, $\mathbf{r}$, as indicated in panel (b2).
Panel (d) shows the phase diagram $V\times \theta$ for fixed $U$ and $\varphi$: data points are exact diagonalization results (full lines guide the eye) for $t=1$, while the dashed line corresponds to the atomic limit ($t=0$). 
}
\label{fig:corr-tp}
\end{figure*}

\paragraph{Model and methods.---}With this in mind, here we consider \textit{spinful} atoms (i.e., a mixture of atoms in two hyperfine states) on a half-filled optical lattice.
The system is described by the Hamiltonian, 
\begin{align}
\mathcal{H} =&-t\sum_{\langle \mathbf{i,j}\rangle\!,\sigma}\left(c_{\mathbf{i}\sigma}^\dagger c_{\mathbf{j}\sigma}^{\phantom{\dagger}}+\mathrm{H.c.}\right)
+U\sum_{\mathbf{i}}n_{\mathbf{i}\uparrow}n_{\mathbf{i}\downarrow}\nonumber\\
&+\sum_{\mathbf{i}\neq\mathbf{j}} V_{\mathbf{i}\mathbf{j}}\ n_\mathbf{i}\,n_{\mathbf{j}},
\label{eq:dEHM}
\end{align}
where, $c_{\mathbf{i}\sigma}$ ($c^\dagger_{\mathbf{i}\sigma}$) denotes the particle annihilation (creation) operator and $n_\mathbf{i}$ the number operator at site $\bf i$.  The sums run over sites of a square optical lattice, with $\langle \mathbf{i,j}\rangle$ denoting nearest neighbor sites; $\sigma=\,\uparrow,\downarrow$ denotes the two hyperfine states, and $t$ is the hopping integral. 
An external field aligns the dipoles parallel to the unit vector $\hat{\mathbf{d}}$, specified by the usual polar angles $\theta$ and $\varphi$, 
taking $\hat{\mathbf{z}}$ perpendicular to the square lattice; see Fig.\,\ref{fig:corr-tp}(d).
The dipolar interaction is then written as
\begin{equation}
	V_{\mathbf{i}\mathbf{j}}=\frac{V}{r_\mathbf{ij}^3}\left[1-3\left(\hat{\mathbf{r}}\cdot\hat{\mathbf{d}}\right)^{2}\right],
\end{equation}
where $V$ (proportional to the square of the dipole moments) is the strength of the interaction, $\mathbf{r_{ij}}\equiv\mathbf{i}-\mathbf{j}$ is a vector joining sites on the lattice, and $\hat{\mathbf{r}}$ is its unit vector. 
The interaction of two atoms in the same optical well, $U$, is the sum of two contributions \cite{Goral02,Lahaye09}: one is the usual contact interaction, tunable through a Feshbach resonance; the other comes from the dipolar interaction, whose behavior at small distances is limited by the finite size of the atoms \cite{Goral02,Lahaye09}.

The ground state properties of the aforementioned Hamiltonian is analysed with the Lanczos method  on a $4\times 4$ lattice with periodic boundary conditions, in the subspace of half filling; translational symmetry and total spin projection are also incorporated in the bases used \cite{Roomany80,Gagliano86}.
In line with experiments in the absence of dipolar interactions, here we consider the case $U=8t$, which is also convenient since finite-size effects are small in the strong-coupling regime -- more on this below. 
The finite lattice size we use forces us to truncate the dipolar interaction beyond second neighbors. 
Nonetheless, anisotropy and competition between attractive and repulsive couplings are preserved.
We also perform strong-coupling analyses, complemented by simulated annealing, in order to check the consistency of exact diagonalization results and to consider
the effects of thermal fluctuations.

Here we borrow the attributes \emph{spin} and \emph{charge,} familiar from the condensed matter context, to respectively denote \emph{atomic species} and \emph{atomic site density}.  
Accordingly, in terms of $\hat{m}_\mathbf{r}=\hat{n}_{\mathbf{r}\uparrow}-\hat{n}_{\mathbf{r}\downarrow}$  and $\hat{n}_\mathbf{r}=\hat{n}_{\mathbf{r}\uparrow}+\hat{n}_{\mathbf{r}\downarrow}$ we define the following correlation functions: 
spin-spin, $C_s(\mathbf{r})\equiv\langle\hat{m}_\mathbf{0} \hat{m}_\mathbf{r}\rangle$,
charge-charge, $C_c(\mathbf{r})\equiv\langle\hat{n}_\mathbf{0} \hat{n}_\mathbf{r}\rangle$,
and local moment-local moment (from now on referred to as moment-moment), $C_m(\mathbf{r})\equiv\langle\hat{m}^2_\mathbf{0} \hat{m}^2_\mathbf{r}\rangle-\langle\hat{m}^2_\mathbf{0}\rangle\langle \hat{m}^2_\mathbf{r}\rangle$; 
this latter quantity is most readily accessible in experiments \cite{Cheuk15,Parsons16,Boll16,Cheuk16}, and, as we will see, carries the signature of both quantum and thermal phase transitions. 

\begin{figure*}[t]
\includegraphics[scale=0.36]{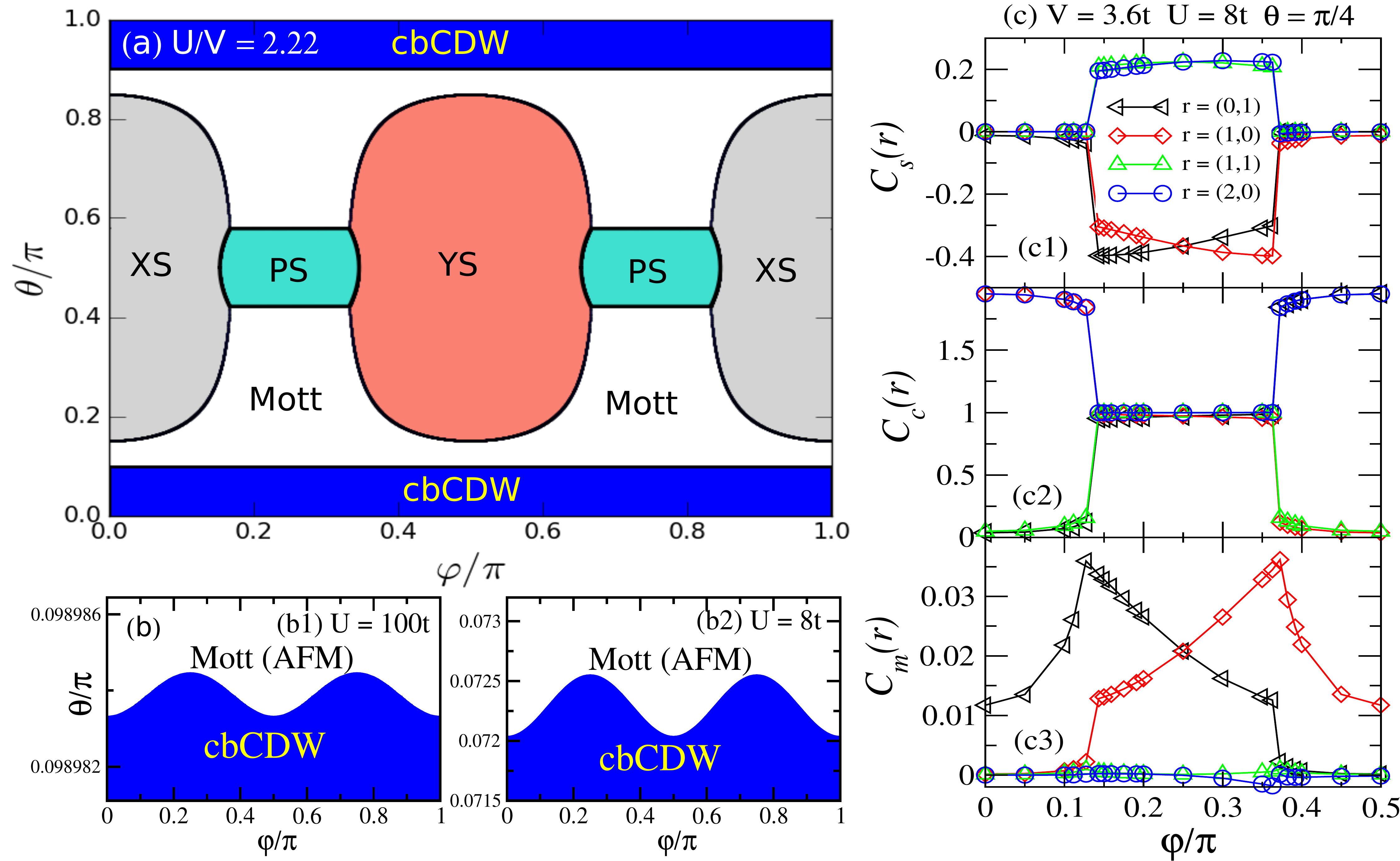} 
\caption{(Color online) Ground state phase diagram in terms of $\theta$ and $\varphi$, in the atomic limit for $U/V=2.22$ (a), striped phases along X (XS) and Y (YS) directions, phase separated phases (PS), Mott insulating phase, and checkerboard CDW  (cbCDW) phases are present.  Second order perturbation theory results for $U/t=100$ (b1) and $U/t=8$ (b2) and Lanczos data for the dependence of different correlation functions with $\varphi$  for $U=8t$, $V=3.6t$ and $\theta=\pi/4$ (c).  
}
\label{fig:ph_diag-th_phi}
\end{figure*}

\paragraph{Zero-temperature transitions.---}Let us first fix the direction of polarization and vary the strength of the dipolar interaction, $V$. 
Figures \ref{fig:corr-tp}(a) show the correlation functions in the isotropic case, $\theta\!\!=\!\!\varphi\!\!=\!\!0$: spin correlations consistent with a N\'eel-like arrangement (a1) are completely suppressed at $V_{\rm CB}\approx 3.1$, beyond which charge correlations (a2) develop. 
The system therefore goes from a Mott phase, in which each species occupies one sublattice, to a checkerboard charge density wave (cbCDW) phase, in which only one of the sublattices is occupied by both species; see cartoons in Fig.\,\ref{fig:corr-tp}(d).
Panel (a3) shows the moment-moment correlation function, which captures the increase of fluctuations at the critical point; the sharp drop in the local moment is responsible for the sharpness of $C_m(\mathbf{r})$  at the transition.
By contrast, when the dipoles point along the $\hat{\mathbf{x}}$ direction [Figs.\,\ref{fig:corr-tp}(b)] the transition is from a Mott phase to a striped phase, at a smaller $V_c$ than for the isotropic case; the direction of the stripes is that of the dipoles, 
since arranging them head-to-tail lowers the energy and skipping a row costs less energy than placing them on adjacent rows. 
As a result, nearest-neighbor spin correlations are now anisotropic in the Mott phase:
in strong coupling, $J_{\nu,\mathrm{eff}}=4t^2/(U-V_\nu)$, $\nu=x,y$, with $V_x<0$ and $V_y>0$, so that attraction weakens magnetic correlations. 
By the same token, local moment fluctuations are also anisotropic, since vertically one has doublon-holon pairs while horizontally one has doublon-doublon pairs, the latter being less prone to fluctuations than the former.

Since the nature of the CDW state depends on the polarization angle, we now probe the phase transitions driven by changing the direction of the dipoles within the $xz$-plane ($\varphi=0$), while $V$ is kept fixed. 
Figures~\ref{fig:corr-tp}(c) show that with increasing $\theta$ the cbCDW phase gives way to a Mott phase (with anisotropic correlations), and further increase in $\theta$ drives the system to another CDW phase, now with stripes along the $\hat{\mathbf{x}}$ direction (XS); see the dash-dotted line in Fig.\,\ref{fig:corr-tp}(d).
This intervening Mott phase disappears at some critical $V_c$, which is not very sensitive to the presence of the hopping for a fixed $U/V$ in the physically relevant domain of $U\gg t$, 
as revealed by comparing with the size-independent strong coupling results; see Fig.\,\ref{fig:corr-tp}(d).
Note that for $V< V_\mathrm{CB}\approx 3.1t$ no cbCDW state is formed, and 
the smaller $V$ gets, the direction of polarization must get closer to the plane in order to reach the XS phase; interestingly, below $V_\textrm{H}\approx 1.5t$ no CDW is formed.

In order to relax the constraint of polarization within the $xz$-plane, we have taken advantage of the fact that the atomic limit (i.e., $t\to0$) captures, to a very good approximation, the essence of the phase diagrams, as discussed in connection with Fig.\,\ref{fig:corr-tp}(d). 
Accordingly, we have mapped out the lowest energy states in the thermodynamic limit at fixed $U$ and $V$, for many values of $\theta$ and $\varphi$; consistency with data from both Lanczos diagonalizations and simulated annealing was checked in many cases.  
Our findings can be summarized in the $\theta\times\varphi$ phase diagram of Fig.\,\ref{fig:ph_diag-th_phi}(a), for $U/V=2.22$, which displays the symmetry under a reflection of the polarization with respect to the plane of the lattice.
For polarization nearly perpendicular to the plane ($ 0 \leq \theta\lesssim 0.1\pi $ and  $ 0.9 \pi \lesssim \theta\leq \pi $), we see that the cbCDW pattern is robust against any rotation of $ \hat{\mathbf{d}} $ around the $z$-axis. 
Figures \ref{fig:ph_diag-th_phi}(b1)  
and \ref{fig:ph_diag-th_phi}(b2) 
show  results from perturbation theory~\cite{SM} indicating that the effect of a finite hopping is to introduce oscillations of negligible amplitudes on the border between cbCDW and Mott phases.

Beyond $\theta \simeq 0.1\pi$, the patterns formed depend on $\theta$ and $\varphi$. 
First, striped phases emerge along either the $\hat{\mathbf{x}}$ direction (XS) for $0\leq\varphi\lesssim 0.2\pi$ (and $0.8\pi\lesssim\varphi \leq \pi$), or the $\hat{\mathbf{y}}$ direction (YS) for  $0.3\pi\lesssim\varphi\lesssim 0.7\pi$.   
Figure \ref{fig:ph_diag-th_phi}(c) shows correlation functions for $\theta= \pi/4$ obtained by means of Lanczos diagonalization; the ground state goes from the XS to the Mott phase around $\varphi/\pi =0.12$ and then to YS phase at $\varphi/\pi=0.38$.
Figure \ref{fig:ph_diag-th_phi}(c3) shows that the spatial anisotropy between doublon-holon and doublon-doublon correlations is picked up by the moment-moment correlation functions as the polarization rotates around $\hat{\mathbf{z}}$, thus confirming its important role in probing quantum phase transitions.

Second, for nearly in-plane polarization  ($0.4\pi\lesssim\varphi\lesssim 0.6\pi$), phase separation (PS) sets in between the XS and YS phases \cite{SM}.
We note that as $U/V$ increases, first the cbCDW phase disappears (for $U/V > 2.586$); then, for $U/V > 2.71$ the PS states are suppressed (see Fig.S2 in Supplemental Material). 
And, finally, for $U/V > 5.33$ the striped phases disappear; in this latter regime, the system is in a Mott state for all polarization directions.

\begin{figure}[t]
\includegraphics[width=0.47\textwidth]{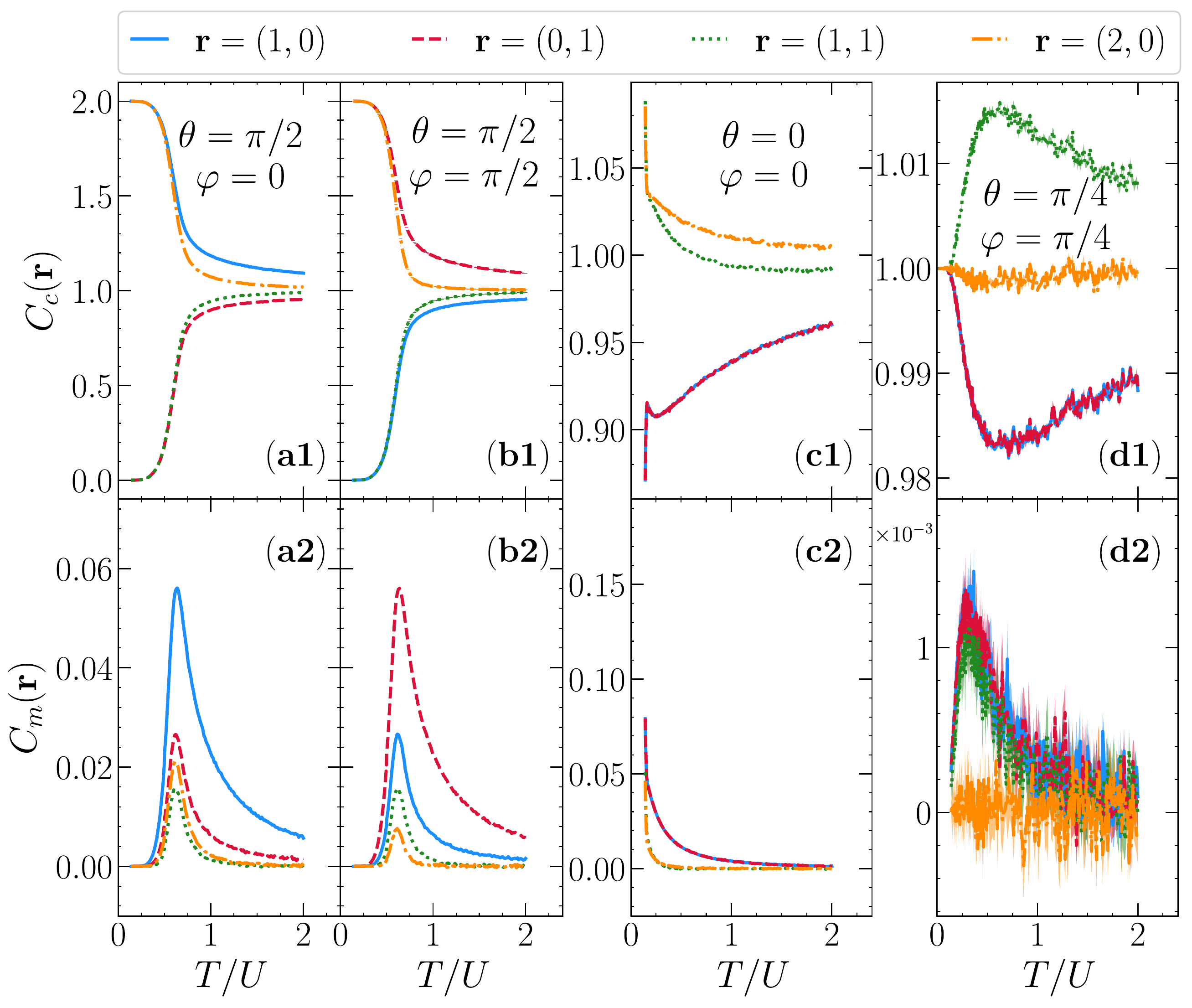} 
\caption{(Color online) Parallel tempering data for charge-charge (top panels) and moment-moment (bottom panels) correlation functions as a function of temperature for (a) $\theta= \pi/2$ and $\varphi=0$ (XS phase), (b)  $\theta= \varphi= \pi/2$  (YS phase), (c)  $\theta= \varphi= 0$ (cbCDW phase) and (d) $\theta= \varphi= \pi/4$ (Mott phase). Data are for $8 \times 8 $ lattices with $U=8$ and $U/V = 2.22$.
}
\label{fig:parallel-temp}
\end{figure}

\paragraph{Thermal transitions.---} Having characterized the ground-state phases and its transitions in terms of the dipole orientations and magnitude of the interactions, an important question, even more prominently from an experimental standpoint, refers to the robustness of these phases in the presence of thermal fluctuations. 
Figure~\ref{fig:parallel-temp} shows the temperature dependence of the charge-charge and moment-moment correlation functions for different directions of polarization, obtained through parallel tempering simulations in the atomic limit~\cite{SM}.
As expected, the charge correlations start at their ground-state values consistent with XS and YS phases [Figs.\,\ref{fig:parallel-temp}(a1) and (b1), respectively], and decrease in magnitude as $T$ increases. 
An estimate of the temperature scale marking the suppression of these ordered phases can be obtained from the peak position of the moment-moment correlations, shown in Figs.\,\ref{fig:parallel-temp}(a2) and (b2): they are the same for both XS and YS phases, namely $T_{\rm XS}/U=T_{\rm YS}/U\simeq0.61$, for $U/V=2.22$. 
For these values of $U$ and $V$, we estimate from Figs.\,\ref{fig:parallel-temp}(c1) and (c2) the ordering temperature for the cbCDW phase as $T/U\sim0.1$, which lies in a range in which parallel tempering simulations are hindered by trapped metastable configurations.
Nonetheless, we are able to infer an upper bound $T_{\rm cbCDW}  <  T_{\rm XS}$, which is valid for different values of the ratio $U/V (<5.33)$. 
One can understand this result by noticing that charge gaps are larger for the striped phases  than for the cbCDW phase, thus leading to higher critical temperatures. 
Finally, for polarizations leading to the Mott phase, such as $\theta=\varphi=\pi/4$ shown in Fig.\,\ref{fig:parallel-temp}(d1) and (d2), the atomic limit also displays a critical temperature associated with the onset of a homogeneous charge ordering, though without any manifest spin order, which is absent in this regime due to the vanishing exchange couplings when $t \to 0$.

\begin{figure}[t]
\includegraphics[width=0.4\textwidth]{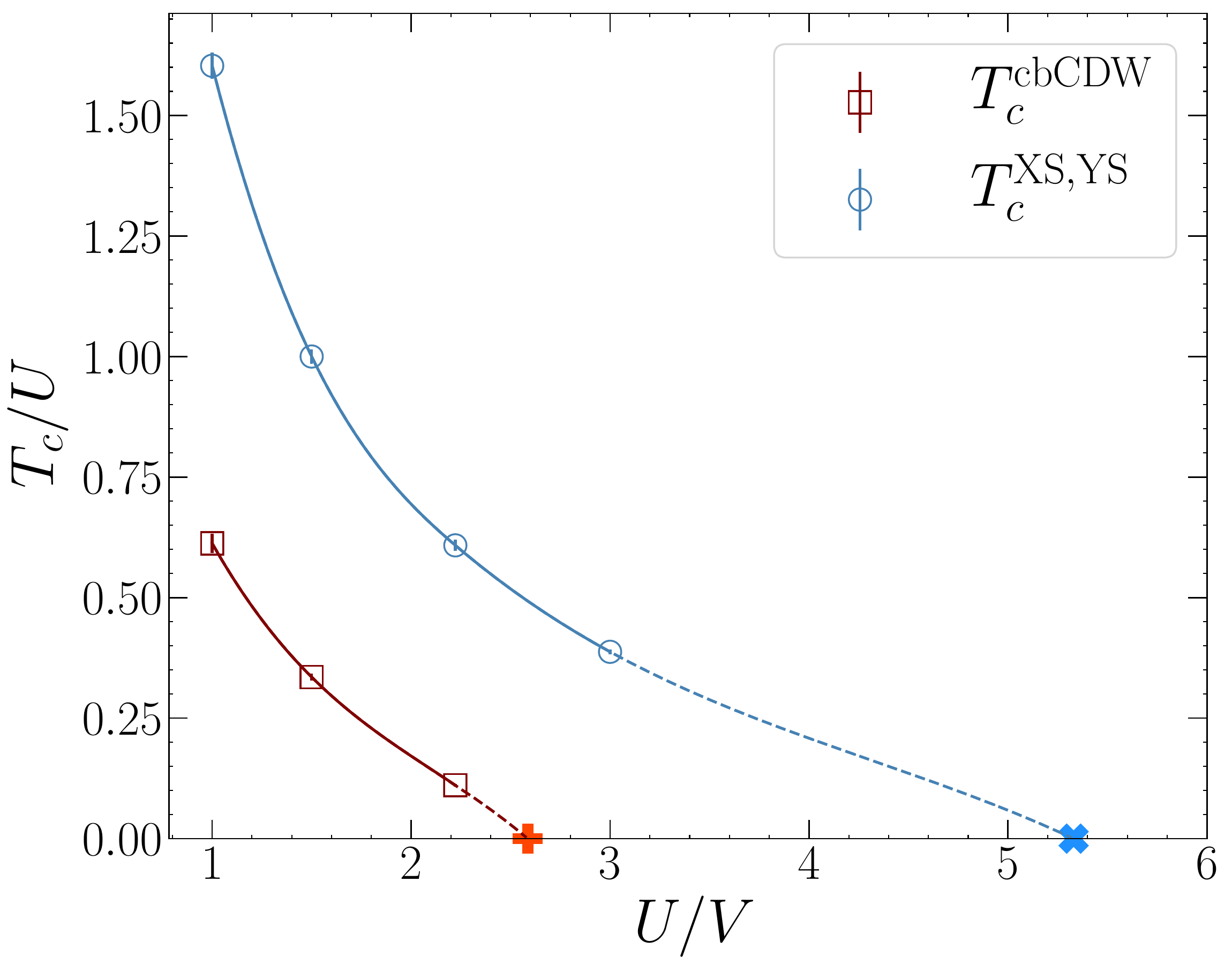} 
\caption{(Color online) Critical temperature for the XS ($\theta=\pi/2$ $\varphi=0$), YS ($\theta=\varphi= \pi/2$) and cbCDW ($\theta=\varphi=0$) phases as a function of $U/V$. Filled markers at $T = 0$ denote the atomic limit results when $L\to\infty$~\cite{SM}, associated to the onset of the Mott insulating phase for the corresponding dipole orientations.
}
\label{fig:tc-vs-UV}
\end{figure}

The estimates thus obtained for the ordering of the XS, YS and cbCDW phases are gathered in Fig.~\ref{fig:tc-vs-UV}; we recall that for $U/V > 5.33$ the ground state is a Mott `insulator' for all polarization directions. 
Recent experiments have reached temperatures as low as $T/U \sim 0.4$ \cite{Baier18}, but with a ratio $U/V$ too large ($\sim 50$) to probe the charge ordered phases (see Fig.~\ref{fig:tc-vs-UV}). 
According to our estimates, at this $T/U$ the striped phases are accessible for $U/V \lesssim 3$, and the cbCDW phase for $U/V \lesssim 1.5$.

\paragraph{Summary.---}We have established that dipolar fermionic atoms in an optical lattice provide a setup in which Mott and density-wave states can in principle be stabilized by a simple control of the direction of polarization.  
These density-wave states may be anisotropic (stripe-like) or occupy one of the sublattices; in addition, one may also find anisotropic phase-separated phases.
Depending on the strength of the dipolar interaction, a rotation of the polarization around the $\hat{\mathbf{z}}$ axis can switch between the density-wave states through a succession of phase separated states.  
Our results are based on exact diagonalizations of a dipolar Fermi-Hubbard Hamiltonian on a $4\times 4$ lattice at half filling, in the regime of strong on-site repulsion. 
In this regime, finite-size effects are not too drastic, as evidenced by the comparison with predictions obtained in the atomic limit (hopping $t\to0$), aided by simulated annealing.
By now the use of moment-moment correlations has proven to be a powerful tool to  probe different phases in experiments with ultracold atoms \cite{Cheuk15,Parsons16,Boll16,Cheuk16}, so that our theoretical predictions for this quantity should provide guidance in the experimental search for these phases.
Indeed, despite the low temperatures  achieved in recent experiments, the large $U/V \sim 50$  \cite{Baier18} regime prevented this kaleidoscope of phases from being accessible. 
If experiments were able to reduce $U/V \lesssim 3$, our parallel tempering simulations predict that for 
$T/U \sim 0.4$ the striped phases will be within reach.


\begin{acknowledgments}
TMS, TP and RRdS acknowledge support by the Brazilian Agencies 
CNPq, CAPES, FAPERJ and INCT on Quantum Information; they are also grateful to CSRC for the hospitality while this work was concluded. RM acknowledges support from NSAF-U1530401 and from the National Natural Science Foundation of China (NSFC) Grant No. 11674021  and No. 11650110441.  The computations were performed in the Tianhe-2JK at the Beijing Computational Science Research Center (CSRC).
\end{acknowledgments}



\bibliography{dhm.bib}

\newpage

\widetext
\clearpage

\onecolumngrid
\begin{center}
  \textbf{\large Supplementary Material: \\ \medskip
\large A kaleidoscope of phases in the dipolar Hubbard model}\\[.2cm]
\end{center}

\twocolumngrid

\setcounter{equation}{0}
\setcounter{figure}{0}
\setcounter{table}{0}
\setcounter{page}{1}
\renewcommand{\theequation}{S\arabic{equation}}
\renewcommand{\thefigure}{S\arabic{figure}}




\section{Atomic limit ($t=0$) phase diagram}

Here we discuss the phase diagram of the dipolar Hubbard model (dHM) in the atomic limit ($t=0$),
\begin{align}
\mathcal{H}_\mathrm{at}= U\sum_{\mathbf{i}}n_{\mathbf{i}\uparrow}n_{\mathbf{i}\downarrow}
+\sum_{\mathbf{i}\neq\mathbf{j}} V_{\mathbf{i}\mathbf{j}}\ n_\mathbf{i}\,n_{\mathbf{j}}.
\label{eq:atomic}
\end{align}
As mentioned in the main text, up to next-nearest neighbors $V_{\mathbf{i}\mathbf{j}}$ becomes 
\begin{align}
	&V_x\equiv V\left(1-3\sin^2\theta\cos^2\varphi\right)\\
	&V_y\equiv V\left(1-3\sin^2\theta\sin^2\varphi\right)\\
	&V_{d1}\equiv\frac{V}{2^{3/2}}\left[1-\frac{3}{2}\sin^2\theta\left(1+\sin2\varphi\right)\right]\\
	&V_{d2}\equiv\frac{V}{2^{3/2}}\left[1-\frac{3}{2}\sin^2\theta\left(\cos2\varphi+\sin2\varphi\right)\right]
\end{align}
	
\begin{figure}[b]
\includegraphics[scale=0.35]{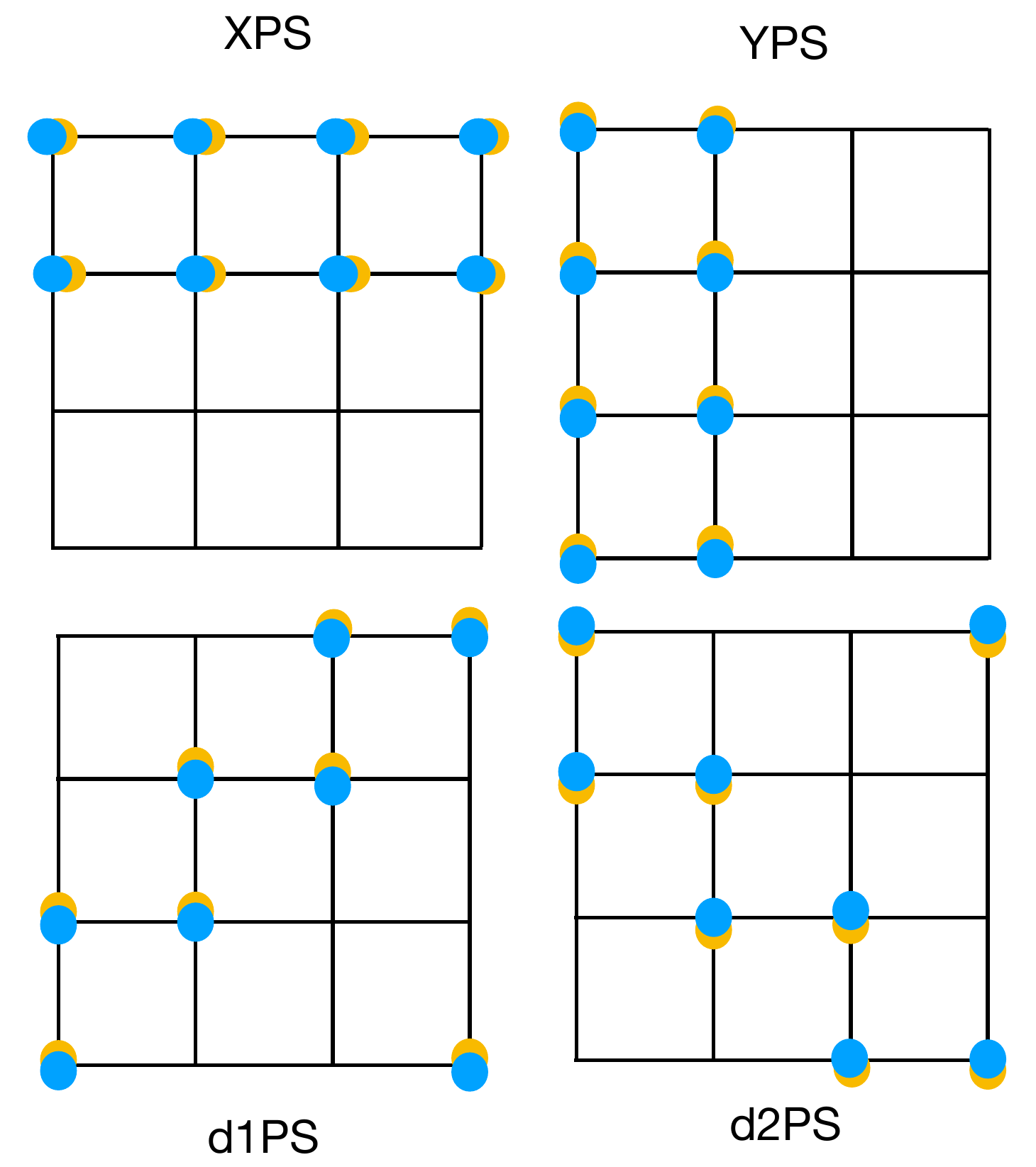} 
\caption{(Color online) Some phase-separated (PS) configurations for a $4\times4$ lattice.}
\label{fig:PhaseSep-cartoons}
\end{figure}

The eigenstates of Eq.\eqref{eq:atomic} are product states (classical states) and the ground state (GS) is the one which minimizes the energy for the given values of $U$, $V$, $\theta$ and $\varphi$.
For instance, when $U \gg V$, double occupancies are suppressed due to the high energy penalty $U$, and the GS corresponds to a Mott insulator.
Physical intuition can be used to set up other possible GS classical states; see, e.g., Fig.\,1 of the main text and Fig.\,\ref{fig:PhaseSep-cartoons}.

For an $L\times L$ lattice with periodic boundary condition, the energy per particle at half filling for the competing ground states may be written as 
\begin{subequations}
\label{eq:E0}
\begin{align}
	\frac{E_0^\mathrm{cbCDW}}{L^2}&=\frac{U}{2}+2(V_{d1}+V_{d_2})\\
	\frac{E_0^\mathrm{Mott}}{L^2}&=V_x+V_y+V_{d1}+V_{d_2}\\
	\frac{E_0^\mathrm{XS}}{L^2}&=\frac{U}{2}+2V_x\\
	\frac{E_0^\mathrm{YS}}{L^2}&=\frac{U}{2}+2V_y\\
	\frac{E_0^\mathrm{XPS}}{L^2}&=\frac{U}{2}+2(V_x+V_y+V_{d1}+V_{d_2})\\
		&-\frac{1}{L}4(V_y+V_{d1}+V_{d2})\\
	\frac{E_0^\mathrm{YPS}}{L^2}&=\frac{U}{2}+2(V_x+V_y+V_{d1}+V_{d_2})\\
		&-\frac{1}{L}4(V_x+V_{d1}+V_{d2})\\
	\frac{E_0^\mathrm{d1PS}}{L^2}&=\frac{U}{2}+2(V_x+V_y+V_{d1}+V_{d_2})\\
		&-\frac{1}{L}4(V_x+V_{y}+2V_{d2})\\
	\frac{E_0^\mathrm{d2PS}}{L^2}&=\frac{U}{2}+2(V_x+V_y+V_{d1}+V_{d_2})\\
		&-\frac{1}{L}4(V_x+V_{y}+2V_{d1}),
\end{align}
\end{subequations}
where XPS and YPS denote $\hat{\mathbf{x}}$- and $\hat{\mathbf{y}}$-oriented phase-separated states, while d1PS and d2PS denote $\pm\pi/4$-oriented phase-separated states; see Fig.\,\ref{fig:PhaseSep-cartoons}.
	
By comparing the energy of these different classical states one is able to draw the atomic-limit $\theta \times \varphi$ phase diagrams presented in Fig.\,2(a) of the main text. 
From the outset we note that the last term in the energy of all phase-separated (PS) states vanish as $L\to\infty$, so that all PS states become degenerate in the thermodynamic limit.
For $U/V = 2.22$, we identify the following transitions, depending on the values of $\theta$ and $\varphi$: (I) cbCDW-Mott, (II) XS(or YS)-Mott, (III) XS(or YS)-PS and (IV) PS-Mott; see Figs.\,1 and \ref{fig:PhaseSep-cartoons}.
Indeed, starting from the isotropic case, $\theta=0$, when the GS is a cbCDW, there is a transition to a Mott state at a critical $\theta_{c1}$, given by
\begin{equation}
\theta_{c1} = \arcsin{\left[ \pm \sqrt{ \frac{2}{3} + \left(  \frac{\sqrt{2}}{2 - 4\sqrt{2}} \right) \frac{2U}{3V}}  \right]}, 
\end{equation}
where the $\pm$ apply to  $\theta_{c1}>\pi/2$ or $\theta_{c1}<\pi/2$, respectively, with the proviso that the cbCDW phase disappears for $U/V > (4\sqrt{2} - 2)/\sqrt{2} \approx 2.586$, which would lead to a complex $\theta_{c1}$. This point is marked in Fig.\,4 of the main text, at the corresponding onset of the Mott phase at zero temperature.
The fact that $\theta_{c1}$ is independent of $\varphi$ gives rise to the straight horizontal line phase boundaries in Fig.\,2(a) of the main text.

Increasing $\theta$ above $\theta_{c1}$ leads to attractive dipolar interactions along the $\hat{\mathbf{x}}$ (or $\hat{\mathbf{y}}$) direction while still being repulsive along $\hat{\mathbf{y}}$ (or $\hat{\mathbf{x}}$). 
This energetically favors stripes along $\hat{\mathbf{x}}$ (or $\hat{\mathbf{y}}$), which we denote by XS (or YS); their regions of stability in the $\theta\times\varphi$ plane are shown in Fig.\,2(a) of the main text.
In-between the XS and YS phases there is a Mott region, whose boundaries depend on both $\theta$ and $\varphi$, for fixed $U/V$. 
 
When  $\theta\approx \pi/2$, the average dipolar interaction is attractive and  the phase-separated states compete with both XS (YS) and the Mott state.
For $\theta= \pi /2 $, an XS-PS transition takes place at $\varphi_{cx}\approx 0.15\pi $ for 
$ 0 \le \varphi\le \pi/4 $, or at $\varphi_{cy}\approx 0.35\pi $ (YS-PS transition in this case) for $ \pi/4 \le \varphi\le \pi /2$, see Fig.\,2(a) 
The PS state has the global minimum energy within the range $ \varphi_{cx} < \varphi < \varphi_{cy}$,
due to the fact that the components of the dipolar interaction are attractive, $V_x, V_y, V_{d1} < 0$, thus favoring the condensation of the particles.
In this case, the PS competes with the Mott phase as the dipole direction deviates from $\theta=\pi/2$; see Fig.\,2(a).
The Mott-PS transition occurs in a line of the phase diagram whose critical value of $\theta$ is given by 
\begin{equation}
\theta_{c2} = \arcsin{\left[ \pm \sqrt{ \left( \frac{2}{3}  + \frac{U}{3V} \frac{\sqrt{2}}{1+2\sqrt{2}} \right) }  \right]},
\label{eq:thetac2}
\end{equation}
where $\pm$ respectively correspond to the critical $\theta$ for $\theta< \pi/2$ and $\theta>\pi/2$; again, note that $\theta_{c2}$ is independent of $\varphi$,
For $U / V > (1 + 2\sqrt{2})/\sqrt{2} \approx2.71$, Eq.\,\eqref{eq:thetac2} yields $\sin \theta_{c2}>1$, so that the PS phase is suppressed, with the Mott state dominating the whole region $ \varphi_{cx} < \varphi < \varphi_{cy}$ of the phase diagram.


\begin{figure}[t]
\includegraphics[scale=0.235]{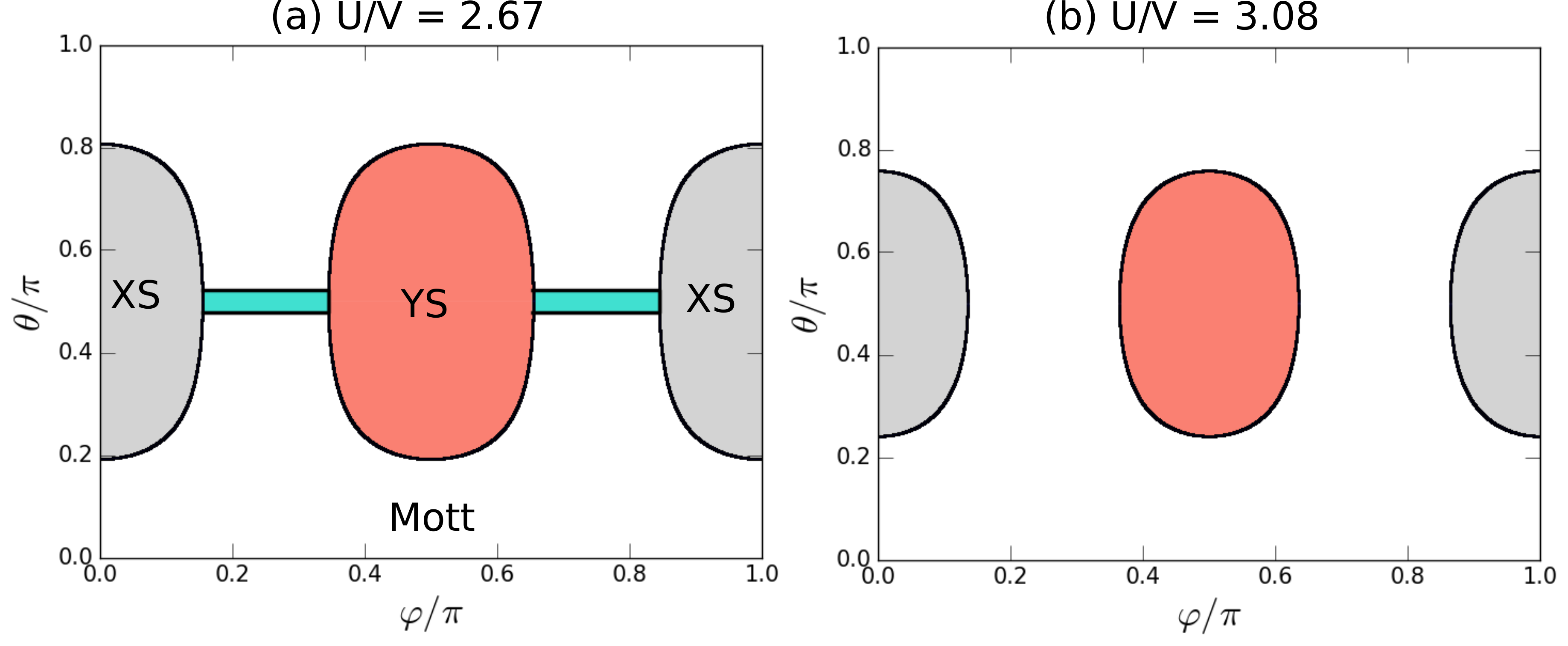} 
\caption{(Color online) As $V$ decreases, first the cbCDW phase is suppressed, then the PS phase is suppressed.}
\label{fig:V3V2_6}
\end{figure}

Summing up, as $U/V$ increases, first the cbCDW phase disappears (for $U/V>2.586$), then for $U/V > 2.71$ the PS states are suppressed; see Fig.\,\ref{fig:V3V2_6}. 
And, finally, for $U/V > 5.33$ the striped phases disappear; in this latter regime, the system is in a Mott state for all polarization directions.

Since experiments with ultra-cold atoms can't always be considered as `in the thermodynamical limit', one must comment on how these results are affected by a finite $L$.
We recall [see Eqs.\,\eqref{eq:E0}] that while the energies per particle for the Mott, XS (or YS), and cb-CDW states are independent of system size, $L$, the PS states have contributions proportional to $1/L$, due to `interface' contributions [see Fig.\,\ref{fig:PhaseSep-cartoons}].
Therefore, in a finite system PS states with different orientations may be formed due to the
anisotropic nature of the dipolar interaction.
For instance, the strip in which the XPS (or YPS) phase is stable when $L\to\infty$ shrinks to a small lobe emerging from the striped phases when, say $L=4$.
By contrast, the boundaries involving non-PS phases are hardly affected by a finite $L$. 

\color{blue}

\color{black}

\section{Second order perturbation theory}

Let us now discuss how a small hopping ($t\ll V, U$) affects the atomic-limit phase diagrams,  resorting to perturbation theory (PT).
The correction to the atomic-limit energies up to second order pertubation theory, $E^{(2)}$, is described by the effective Hamiltonian \cite{Dongen94}
\begin{align}
  \bra{\phi_{0}^{i}}\mathcal{H}_\mathrm{eff}\ket{ \phi_{0}^{j}}= &  \bra{\phi_{0}^{i}}K\ket{\phi_{0}^{j}}  \nonumber\\ 
 & + \sum_{m > 0} \frac{\bra{\phi_{0}^{i}}K\ket{\phi_{m}}\bra{\phi_{m}}K\ket{\phi_{0}^{j}}}{E_{0} - E_{m}},
 \label{eq:sec}
\end{align}
where $E_m$ and $\ket{\phi_m}$  are the respective eigenvalues and eigenstates of $\mathcal{H}_\mathrm{at}$, $\mathcal{H}_\mathrm{at}\ket{\phi_m} = E_m \ket{\phi_m}$.
$\mathcal{H}_\mathrm{eff}$ is therefore an operator which acts in the subspace of
the degenerate ground states of $\mathcal{H}_\mathrm{at}$, $\{\ket{\phi_{0}^{i}}\}$, and the 
pertubation $K$ is the hopping term of the dHM.

The correction $E^{(2)}$ is the lowest energy of $\mathcal{H}_\mathrm{eff}$.
The cb-CDW and the XS(YS) atomic-limit ground states form a subspace that is two-fold degenerate in each case, so $\mathcal{H}_\mathrm{eff}$ is a  $2 \times 2$ diagonal matrix.
In these cases, we obtain  
\begin{align}
\frac{E_\mathrm{cbCDW}^{(2)}}{N} =  & \frac{2 t^{2}}{U -  4 V_{x} - 3 V_{y} + 4 V_{d1} + 4 V_{d2}}  \nonumber\\ 
& + \frac{2 t^{2}}{U -  4 V_{y} - 3 V_{x} + 4 V_{d1} + 4 V_{d2}},
\end{align}
and   
\begin{align}
 \frac{E_\mathrm{XS(YS)}^{(2)}}{N} =   \frac{2t^2}{U +  4 V_{x(y)} - 3 V_{y(x)} - 4 V_{d1} - 4 V_{d2}}.
\end{align}
On the other hand, the atomic-limit Mott states form a  macroscopically degenerate subspace, 
and $\mathcal{H}_\mathrm{eff}$ becomes an anisotropic SU($2$) Heisenberg Hamiltonian \cite{Dongen94,Emery76}
 \begin{equation}
\mathcal{H}_\mathrm{eff} = J_{x} \sum_i \vec{S}_i \cdot \vec{S}_{i\pm{\bf \hat x}} + J_{y} \sum_i \vec{S}_i \cdot \vec{S}_{i\pm{\bf \hat y}} - \frac{N}{4}\left( J_x + J_y \right), 
\end{equation}
where  the exchange couplings $J_x = 4t^2 / (U - V_x)$ and $J_y = 4t^2 / (U - V_y)$ depend on the dipolar angles $\theta$ and $\varphi$. For $J_x$ and $J_y > 0$, the ground state of $\mathcal{H}_\mathrm{eff}$ exhibits an antiferromagnetic order \cite{Sandvik99}.
Here we use linear spin-wave theory \cite{Singh94} to determine the ground state energy of $\mathcal{H}_\mathrm{eff}$, $E_\mathrm{MottAFM}^{(2)}$, for 
different values of $\theta$ and $\varphi$.

\begin{figure}[t]
\includegraphics[scale=0.4]{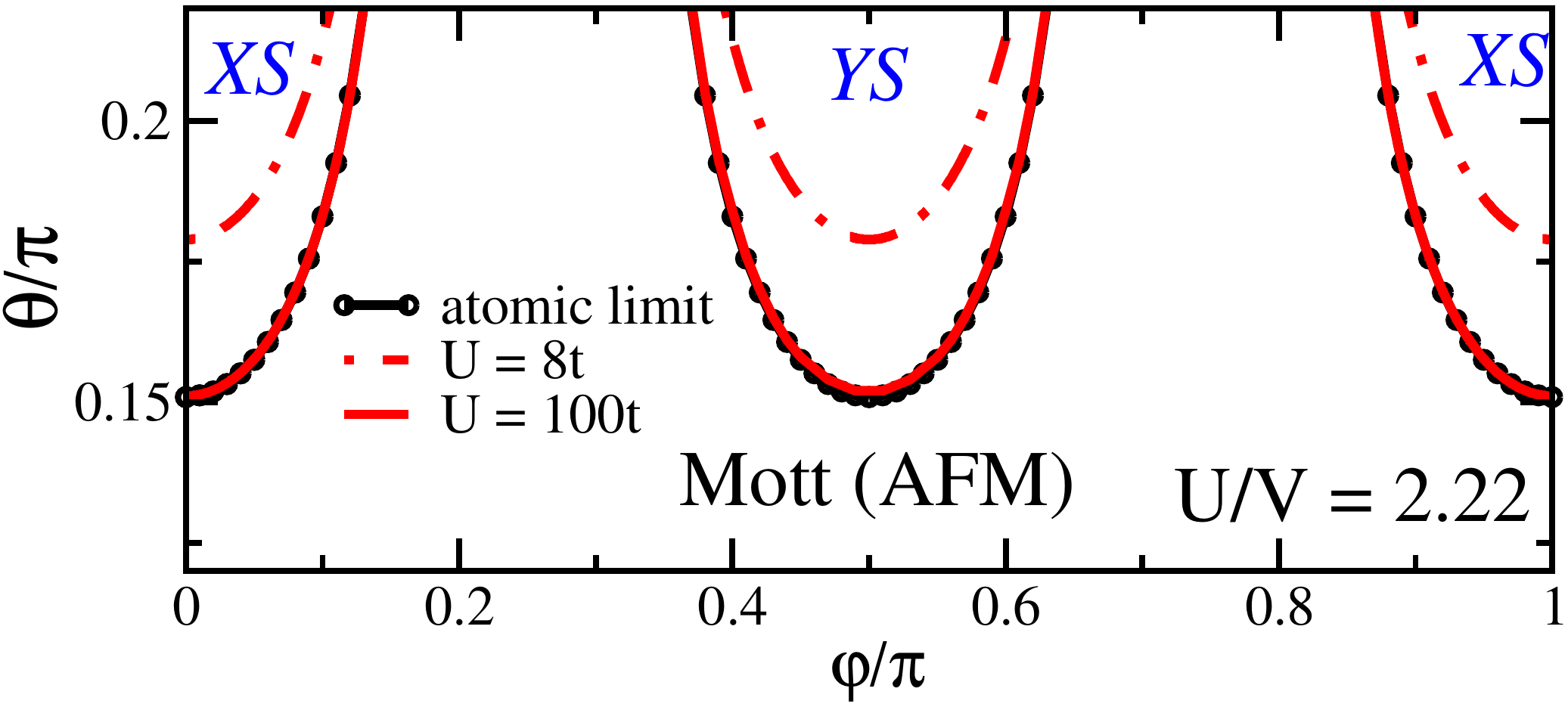} 
\caption{(Color online)  $\theta \times \varphi$ phase diagram  obtained with second order pertubation theory for $U = 100t$ (full red lines) and $U = 8t$ (dashed red lines).
The results for the atomic-limit are also presented. }
\label{fig:phaseSec}
\end{figure}

By comparing the second-order energies, 
\begin{equation}
E_\mathrm{Mott} = E_\mathrm{Mott}^{(0)} + E_\mathrm{Mott}^{(2)},
\end{equation}
\begin{equation}
 E_\mathrm{cbCDW} = E_\mathrm{cbCDW}^{(0)} + E_\mathrm{cbCDW}^{(2)},
\end{equation}
and
\begin{equation}
  E_\mathrm{XS(YS)} = E_\mathrm{XS(YS)}^{(0)} + E_\mathrm{XS(YS)}^{(2)},
\end{equation}
we have established that the main effect of the hopping is to enlarge the region  dominated by the Mott phase in the $\theta \times \varphi$ diagram.
The critical angle $\theta_{c1}$ associated to the Mott-cbCDW transition 
decreases in comparison with the atomic-limit case.
Further, due to the presence of anisotropic AFM correlations, $\theta_{c1}$ acquires a tiny dependence on $\varphi$, as it can be seen from Fig.\,2(b) of the main text.
In addition, the lobes of the $\theta \times \varphi$ phase diagram dominated by the XS and YS phases shrink as we decrease the value of $U$ to $U/t = 8$; see Fig.\,\ref{fig:phaseSec}.


\section{Moment-moment correlations}

As shown in the main text, the moment-moment correlations, $C_m({\bf r})= \left<m^2_{{\bf 0}}m^2_{{\bf r}} \right> - \left< m^2_{{\bf r}}\right>\left< m^2_{{\bf 0}}\right> $, can be used to
identify not just the different transitions described by the dHM, but also the nature 
of the anisotropic CDW phases.
In this section we take a closer look at the local moment, $\langle m^2\rangle$, at the different Mott-CDW transitions discussed in the main text.

\begin{figure}[h]
\includegraphics[scale=0.4]{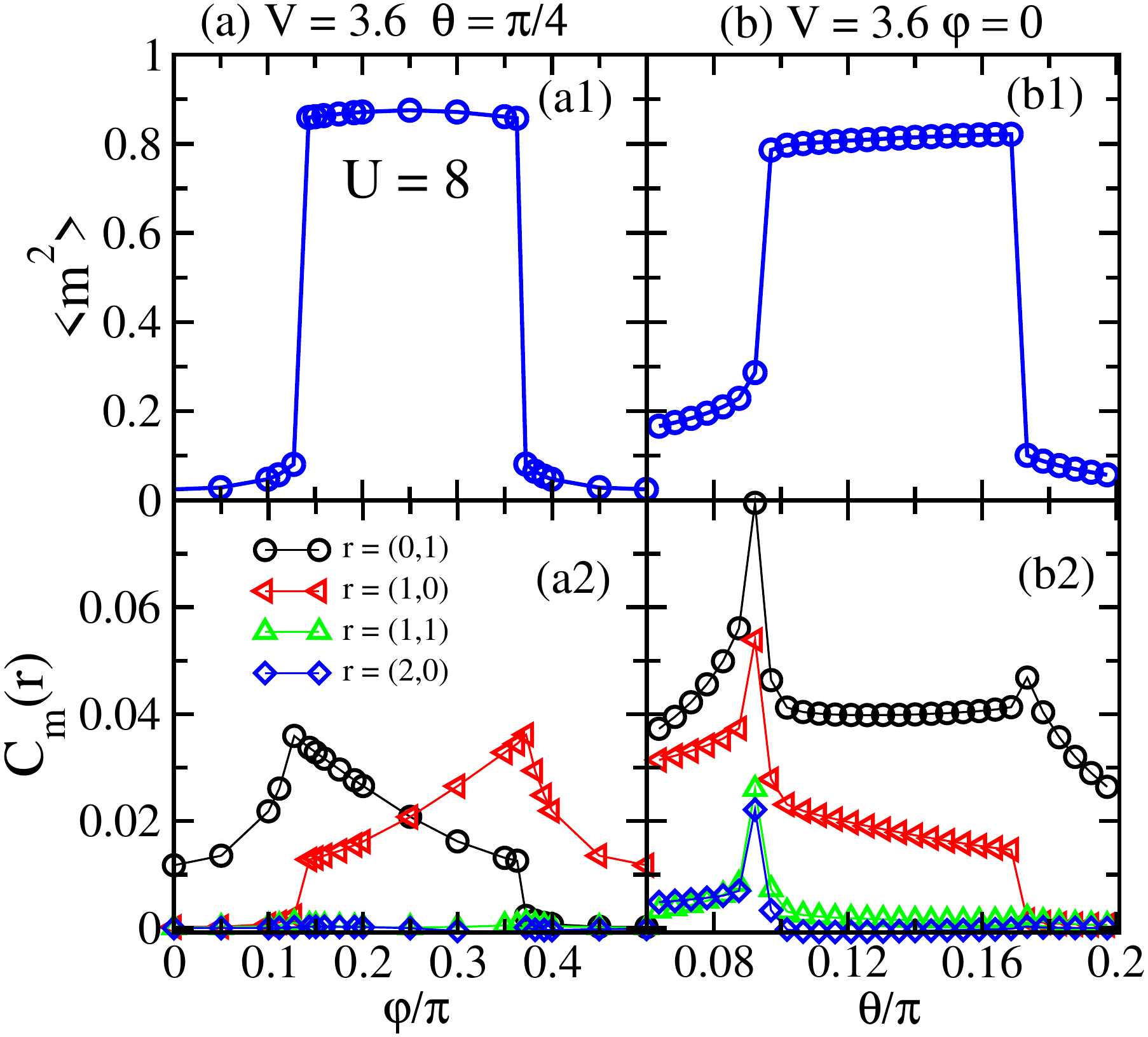} 
\caption{(Color online) Dependence of the local moment and moment-moment correlations with the angle (a) $\varphi$ for  $\theta = \pi/4$  and (b) $\theta$ for $\varphi = 0$.
For both cases we consider $U = 8t$ and $V = 3.6t$}
\label{fig:momentum}
\end{figure}

We first consider the transitions occurring as one varies $\varphi$ for fixed $\theta = \pi/4$;
see  Fig.\,2(c) of the main text.
In this case, Fig.\,\ref{fig:momentum}(a1) shows that $\langle m^2\rangle$ is close to saturation in the Mott phase, but sharply decreases in the striped phases; a similar behavior occurs as $\theta$ varies with fixed $\varphi$, as in Fig.\,\ref{fig:momentum} (b1).
By contrast, Figs.\,1(c) and 2(c) show that $C_m({\bf r})$ is peaked at the different transitions for some specific directions ${\bf r}$. 
For the Mott-XS(YS) transition, for instance, the peak of $C_m({\bf r})$ occurs when
the doublon-holon fluctuations are the strongest; see Fig.\,\ref{fig:momentum}(b1), and the discussion in the main text.
A peak in $C_m({\bf r})$ is also observed at the cbCDW-MottAFM transition, see Fig.\,\ref{fig:momentum}(b2).
Thus, the sharp drop in the local moment is responsible for the sharpness of $C_m(\mathbf{r})$  at the transition.

\section{Parallel tempering in the atomic limit}

To estimate the critical temperatures $T_c^\alpha$ signaling the onset of the different ordered classical phases, $\alpha$, listed in Eq.~\eqref{eq:E0}, we use the parallel tempering~\cite{Earl2005} (or replica exchange method) of the Hamiltonian~\eqref{eq:atomic}. In summary, we use a Monte Carlo (MC) sampling of the occupations of both species \{$\uparrow$ and $\downarrow$\}, promoting random swaps of site occupancies, complemented by random creation and destruction of particles at different temperatures. These moves are implemented as to obey the detailed balance condition, in a particle-hole symmetric version of Eq.~\eqref{eq:atomic}. This guarantees that on average one keeps $\langle n_\uparrow\rangle = \langle n_\downarrow\rangle = 0.5$. After a single MC sweep, an attempt of swap of the configurations related to adjacent temperatures in a given range is induced and accepted with probability
\begin{equation}
p = \min\{1,\exp[-(\beta_i-\beta_j)(E_j-E_i)]\},
 \label{eq:prob_pt}
\end{equation}
where $\beta_i=1/T_i$ is the inverse temperature of a given configuration $i$ whose associated energy for the Hamiltonian~\eqref{eq:atomic} is $E_i$. 

We typically use square lattices up to $L=32$, and 20,000 MC sweeps, with approximately 300 different temperatures chosen in a way to ensure that the range encompasses the associated critical temperatures $T_c^\alpha$. 
It is a known difficulty of the parallel tempering scheme on how to choose the optimal set of temperatures~\cite{Earl2004,Kone2005} which overcome the trapping of metastable configurations when $T\to 0$. 
Although sub-optimal, we used a simple approach of evenly spaced ones, which is more than sufficient to resolve the critical temperatures associated to the onset of the different phases. 

\begin{figure}[h]
\includegraphics[width=0.98\columnwidth]{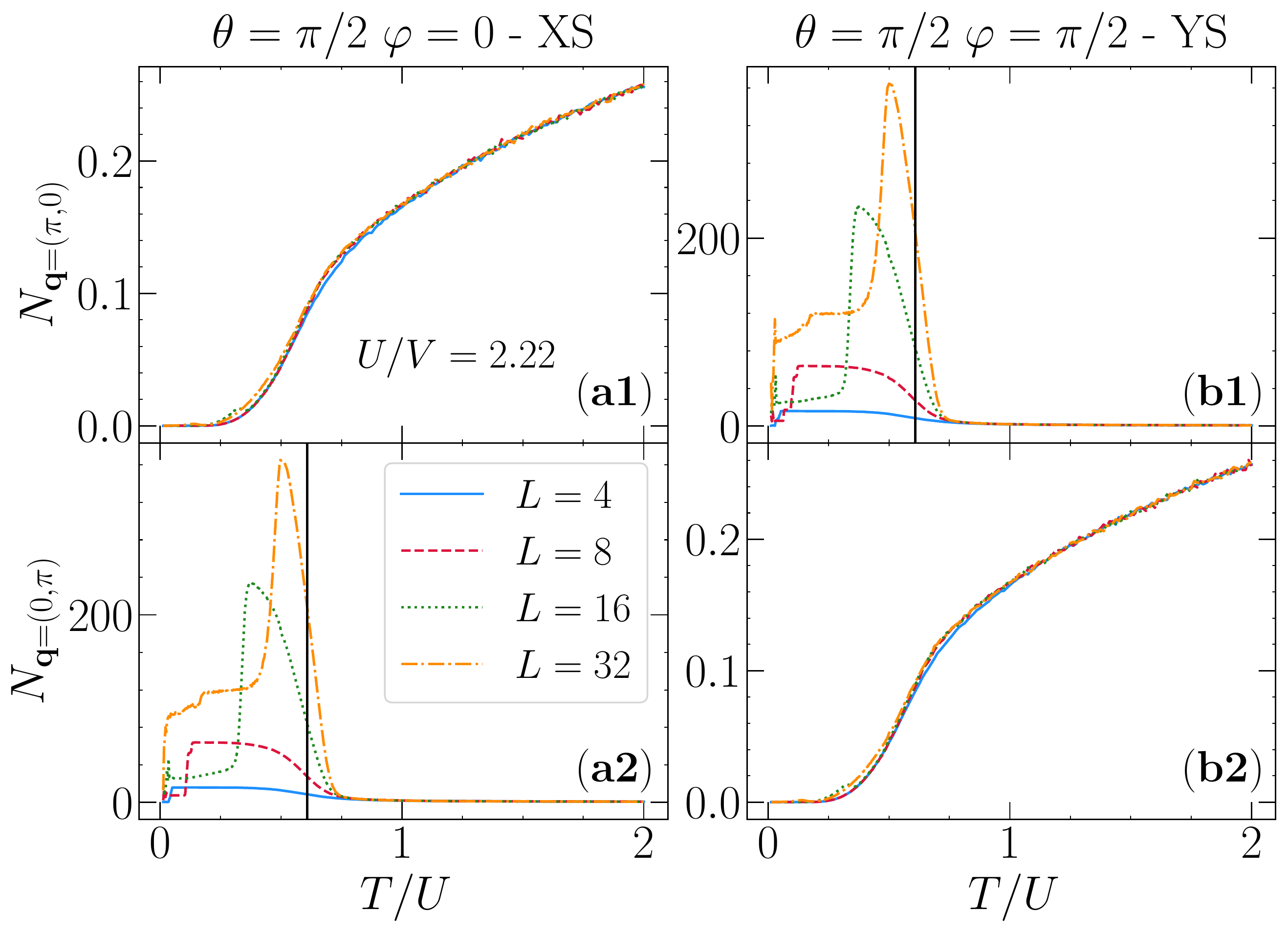} 
\caption{(Color online) Temperature dependence of the charge structure factor for the stripe phases XS in (a) and YS in (b). We select two channels with $\mathbf{q}=(\pi,0)$ in (a1) and (b1), whereas (a2) and (b2) display the $\mathbf{q}=(0, \pi)$ results. Vertical lines depict the peak position of the temperature dependent moment-moment correlation functions, signaling the thermal transition. We fix the interaction ratio $U/V$ to 2.22.}
\label{fig:SF_stripe}
\end{figure}

\begin{figure}[h]
\includegraphics[width=0.75\columnwidth]{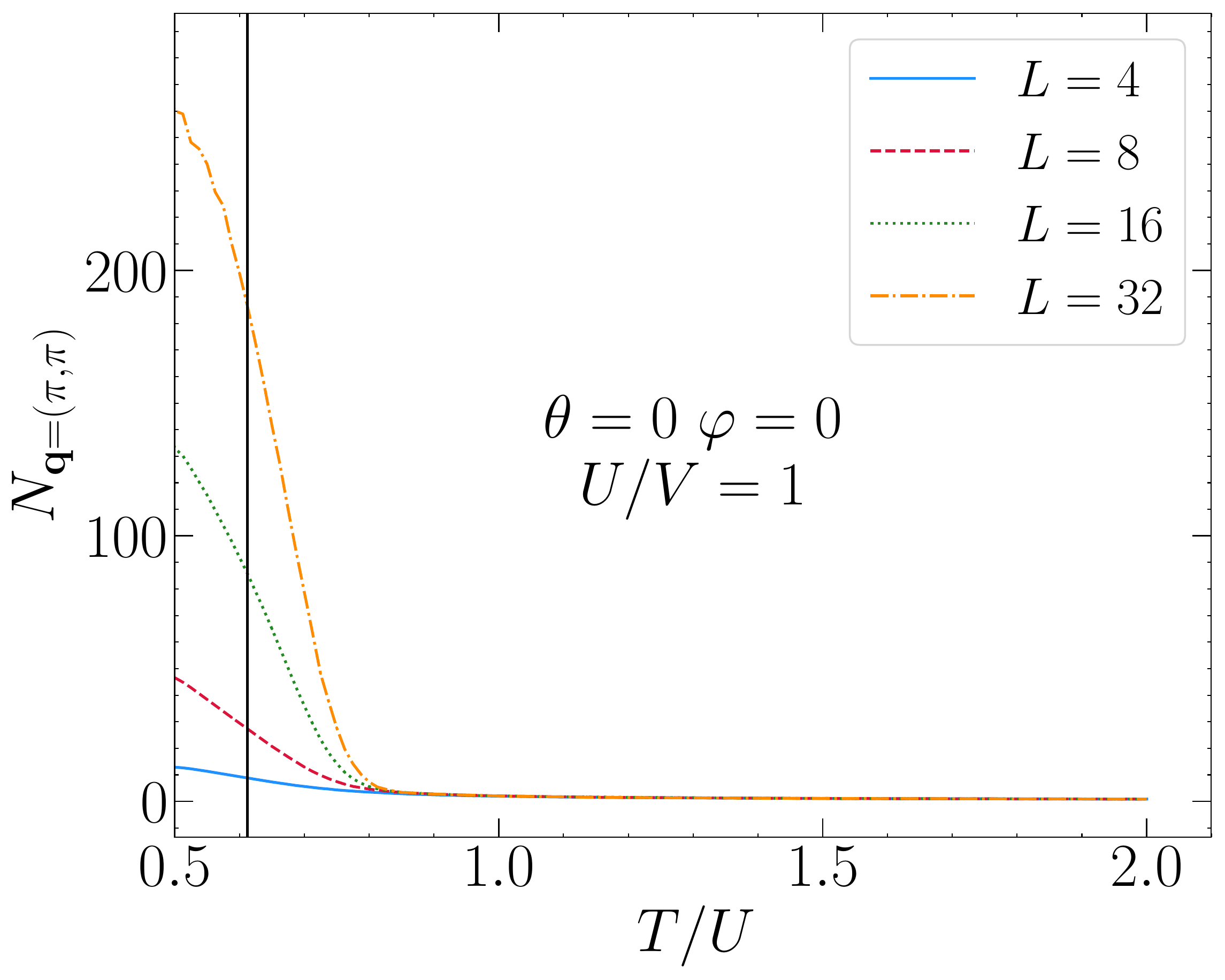} 
\caption{(Color online) Temperature dependence of the ${\bf q} = (\pi,\pi)$ charge structure factor for $\hat {\bf z}$ oriented dipoles. The extensive behavior at low temperatures signals the onset of the cbCDW order. As before, the vertical line depict the peak position of the temperature dependent moment-moment correlation functions. We choose the ratio of interactions $U/V=1$.}
\label{fig:SF_cbCDW}
\end{figure}

Similarly to the quantum version of the Hamiltonian, in the main text
we present local correlations [$C_c(\bf r)$ and $C_m(\bf r)$] which help to identify the charge distribution in all classical phases. Here, to complement this analysis and describe a fully developed order, we compute the associated charge structure factor,
\begin{equation}
 N_{\bf q} = \frac{1}{L^2}\sum_{\bf i,j}\langle e^{ {\rm i}{\bf q\cdot (i - j)} } n_{\bf i}n_{\bf j}\rangle_{\rm MC},
\end{equation}
which becomes an extensive quantity in the presence of a given charge order with wave-vector $\mathbf{q}$.

As an example, we report in Fig.~\ref{fig:SF_stripe} the comparison of $N_{\bf q}$ for two striped phases, XS ($\theta = \pi/2$ and $\phi=0$) and YS ($\theta = \phi  = \pi/2$) in panels (a) and (b), respectively. We note that the structure factor has a symmetric role for different channels: while for XS the ${\bf q}=(0,\pi)$ channel displays an extensive behavior at low temperatures, ${\bf q}=(\pi, 0)$ reflects this corresponding behavior for the YS phase. 
For very low temperatures, however, the  aforementioned trapping of metastable configurations occurs, preventing the observation of a fully formed plateau; this, in turn, signals that the correlation length for this ordering has reached the linear size of the system. 
Nonetheless, the critical temperatures $T_c^{\rm XS}$ and $T_c^{\rm YS}$ lie way above the temperatures where these problems begin to occur. 
As an estimation, we also display as a vertical line in these panels the thermal peak-positions of the moment-moment correlation functions (as in Fig. 3 of the main text), which are very close to the regime where the curves for different system sizes start displaying an extensive behavior. Conversely, for the channels ${\bf q}=(\pi, 0)$ [Fig.~\ref{fig:SF_stripe}(a1)] and ${\bf q}=(0,\pi)$ [Fig.~\ref{fig:SF_stripe}(b2)] for the XS and YS phases, respectively, $N_{\bf q}$ is approximately  independent of the system size, thus confirming the nature of the charge periodicity.

Lastly, we perform similar parallel tempering simulations for the case of isotropic interactions, i.e., $\theta=\varphi=0$, where the ground state of Eq.~\ref{eq:atomic} displays cbCDW order. Figure~\ref{fig:SF_cbCDW} shows the temperature dependence of the ${\bf q}=(\pi,\pi)$ channel for the charge structure factor: As for the stripe phases, at low temperatures this quantity displays an extensive onset, which is close to the peak position of the corresponding moment-moment correlations, thus signaling the checkerboard nature of the charge distribution.





\end{document}